\documentclass[12pt,preprint]{aastex}
\newcommand{\tnm}{\tablenotemark}
\newcommand{\tnt}{\tablenotetext}

\shorttitle{Properties of M Dwarfs}
\shortauthors{T. N. Gautier et al.}

\begin{document}

\title{Far Infrared Properties of M Dwarfs}

\author{Thomas. N. Gautier III \altaffilmark{1},
G. H. Rieke \altaffilmark{2},
John Stansberry \altaffilmark{2},
Geoffrey C. Bryden \altaffilmark{1},
Karl R. Stapelfeldt \altaffilmark{1},
Michael W. Werner \altaffilmark{1},
Charles A. Beichman \altaffilmark{1},
Christine Chen \altaffilmark{3},
Kate Su \altaffilmark{2},
David Trilling \altaffilmark{2},
Brian M. Patten \altaffilmark{4} and 
Thomas L. Roellig \altaffilmark{5}}

\altaffiltext{1}{Jet Propulsion Laboratory, California 
Institute of Technology, Pasadena, California 91109; Thomas.N.Gautier@jpl.nasa.gov}
\altaffiltext{2}{Steward Observatory, University of Arizona, 933 North Cherry Avenue, 
Tucson, Arizona 85721}
\altaffiltext{3}{National Optical Astronomy Observatories, Tucson, Arizona 85719}
\altaffiltext{4}{Harvard-Smithsonian Center for Astrophysics, 60 Garden St., 
                 Cambridge, Massachusetts 02138}
\altaffiltext{5}{NASA Ames Research Center, MS 245-6, Moffett Field, California 94035}

\begin{abstract}

     We report the mid- and far-infrared properties of nearby M dwarfs. 
{\it Spitzer}/MIPS measurements were obtained for a sample of 62 stars 
at 24 \micron, with subsamples of 41 and 20 stars observed at 70 \micron\ 
and 160 \micron\ respectively. We compare the results with current models 
of M star photospheres and look for indications of circumstellar dust in 
the form of significant deviations of \hbox{K-[24 \micron]} colors and 
70 \micron\ / 24 \micron\ flux ratios from the average M star values. At 
24 \micron, all 62 of the targets were detected; 70 \micron\ detections were 
achieved for 20 targets in the subsample observed; and no detections were 
seen in the 160 \micron\ subsample. No clear far-infrared excesses were detected 
in our sample. The average far infrared excess relative to the photospheric 
emission of the M stars is at least four times smaller than the similar average 
for a sample of solar-type stars. However, this limit allows the average 
fractional infrared luminosity in the M-star sample to be similar to that for 
more massive stars. We have also set low limits ($10^{-4}$ to $10^{-9}$ 
M$_{Earth}$ depending on location) for the maximum mass of dust possible 
around our stars.

\end{abstract}


\section{Introduction}\label{intro}

     The study of dust around main sequence stars addresses intriguing questions 
in basic astrophysics while providing insights into the possible existence 
of planetary systems around these stars. The pioneering results from IRAS on 
infrared excess around stars like Vega \citep{au84} rapidly led to 
the realization that the circumstellar dust responsible for this emission 
could not survive for the lifetime of the star and that it must be 
replenished - from cometary or asteroid-type objects - just as the zodiacal 
cloud in our Solar System is replenished. The maintenance of this dust 
ultimately depends on collisions between asteroid-size objects, yielding 
fragments that collide with each other and eventually grind down to fine 
dust that is heated by the star and appears as an infrared excess. Thus the 
characteristics of circumstellar dust around a main sequence star (referred 
to hereafter as a ``debris disk'') reflect the planet-building and 
planet-destroying processes occurring around that star.

     The {\it Spitzer} Space Telescope \citep{wer04}, building on the 
foundation established by IRAS and ISO, has already produced major advances 
in our understanding of the debris disk phenomenon. These include 
demonstration of the stochastic nature of debris disk evolution 
\citep{gr05,su06,bei05a}, the first detection of debris disks around F, G, and 
K stars with known radial velocity companions \citep{bei05b} and investigations 
of the frequency of occurence of debris disks around solar type stars \citep{kim05,bry05}. Here 
we present the results from a {\it Spitzer} search for dust associated 
with M dwarfs.

     M dwarfs, with masses between 0.08 and 0.5 M$_\odot$, are the most 
common stars in the galactic plane. Thus the study of debris disks around them 
is needed to complete our understanding of the nature and incidence of debris 
disks in the solar neighborhood and the dependence of planet-forming processes 
on the mass of the central star. Since M stars are the most numerous 
stellar type, measurement of their debris disk frequency has large 
implications for the overall population of planetary systems in the galaxy. 
It may be that the nearest extrasolar planets to the Sun are orbiting M 
stars, a possibility that would be supported if M dwarfs frequently host 
debris systems.

     However, the timescales and processes for dust expulsion may be considerably 
different for M stars than for their more luminous cousins \citep{plav05}. 
Also, early results from precision radial velocity surveys have suggested that M stars 
have a lower frequency of giant planets than F and G stars \citep{but04}. It is 
important to determine if such a marked difference is also present in the relative 
frequency of debris disks between the M dwarfs and earlier types.

     Finally, besides extending our understanding of the evolution of planetary 
systems by investigating the frequency of debris disks around M dwarfs, 
these far-infrared measurements can determine basic parameters of the M 
stars themselves. We compare our data with current models of M dwarf 
photospheric emission and demonstrate how to refine determinations of M 
dwarf effective temperatures with the accurate 24 \micron\ data available 
with {\it Spitzer}.

\section{New Observations}\label{obs}

     We observed 62, 41, and 20 M dwarfs, respectively, at 24, 70, and 160$\mu$m, 
using the Multiband Imaging Photometer for {\it Spitzer} \citep[MIPS;][]{gr04}. 
Our target stars are listed in Table \ref{data_table}. Stars of spectral type M7 and 
earlier (a total of 40), plus one M9 star, are part of a MIPS program to 
obtain photometry of all stars lying within 5 pc of the sun ({\it Spitzer} program 
ID 52). Twenty one later-type stars (M6 - M9) lying at distances of 6 to 18 pc 
were generally selected to be the brightest members of their spectral classes 
({\it Spitzer} program ID 56). These later-type stars were observed only at 24 \micron. 
While age determination of field M stars is notoriously difficult it is likely 
that the stars in our sample are nearly all older than 1 Gy.

     The 24 $\mu $m observations were made using the MIPS photometry observation 
template with the small dither size. For the nearby stars we generally used 
1 cycle of the dither pattern (48s integration time). IL Aqr, LHS 3685 and LP 944-20 
were exceptions receiving 180s, 136s and 1500s integration time respectively. This strategy 
typically resulted in very high signal-to-noise ratios (SNRs) at 24 \micron. 
The more distant, later type, stars received from 1 to 9 dither 
cycles (48s to 1380s) at 24 \micron; many of them were measured only to SNR $\sim\! 10$.

     The 70 $\mu $m observations were also made using the MIPS photometry 
observation template with the small dither size and the default image scale 
\citep{gr04}. In this band, the required integration times were much longer 
than at 24 \micron\ and therefore automatically resulted in several cycles of 
the dither pattern. We limited the maximum integration time at 70 \micron\ to 
$\sim\! 1000$ seconds, both to fit the program within its time allocation and 
because the noise behavior of MIPS at longer integration times was unknown at 
the time these observations were scheduled (before the launch of {\it Spitzer}). 
Our goal of SNR $\sim\! 6$ on the photosphere was reached for many of the 
70 \micron\ observations but the limitation on integration time resulted in low 
SNR detections or upper limits at 70 \micron\ for the fainter members of the sample. 

     To explore the possibility that these low-luminosity stars might possess 
very cold debris disks, 160 $\mu $m photometry was carried out for 20 stars 
in the sample. Detection of the stellar photospheres, whose median flux 
densities are expected to be a few mJy, was not achievable given {\it Spitzer's} 
measured extragalactic confusion limit of 8 mJy, 1$\sigma$ \citep{dol04}. 
The observations were therefore designed only to detect a source 3-5 
times the confusion noise level. Four cycles of 160 \micron\ photometry, 
totaling 84 sec of exposure time for each source, were used.


     The data were processed using the MIPS instrument team Data Analysis Tool 
(DAT, version 2.73) described by \citet{gord05}. For the 24 \micron\ data, 
basic processing included slope fitting, flat-fielding, and corrections for 
droop and ``jailbar'' artifacts (DC offsets between adjacent detector columns 
that appear in measurements of bright sources). Further details are given by
\citet{eng06}. The 70 \micron\ data processing also used the DAT, but was similar to 
that of the {\it Spitzer} pipeline version S12. Details are given in 
\citet{gord04}.  Mosaics were constructed using pixels interpolated to about 
1/2 the native pixel scale of the arrays. 

     Fluxes were measured using fairly small apertures to improve the 
signal-to-noise ratios of the measurements, particularly given 
the relatively high spatial variability of the sky background at 70 \micron. 
Most targets were measured using apertures 9.96\arcsec\ and 39.4\arcsec\ 
in diameter at 24 and 70 \micron. The observed PSF full width at half maximum 
is 6.4\arcsec\ and 19.3\arcsec\ , at 24 and 70 \micron\ respectively. In a few 
cases nearby sources contributed significant flux at the target location, so we 
used apertures 25\% smaller than those just described to reduce contamination. 
In one case (LHS 58 \& 59, identified in the notes to Table \ref{data_table}), 
we deblended two sources at 70 \micron\ using the model PSFs. 
The $\sim10$\arcsec\ aperture at 24 \micron\  corresponds to disk radii of  
7-90 AU at the various distances of our stars. At 70 \micron, where only stars 
within 5 pc were observed, the $\sim40$\arcsec\ aperture corresponds to disk radii 
of 36-200 AU.

After extraction, the measured fluxes were corrected to infinite aperture. 
Aperture corrections were computed using STinyTim model PSFs \citep{krist05}
for a 3000 K blackbody source. The model PSFs were smoothed until they provided 
good agreement with observed stellar PSFs out through the first Airy maximum.

     Targets were usually identified in the 24 \micron\ images using their J2000 
coordinates and proper motions from SIMBAD. In most instances the agreement 
between the predicted and observed position was excellent (see Notes to 
Table \ref{data_table} for cases where there was a significant disagreement). 
However, many of our later M type stars are rather dim, even at 24 \micron, and 
special care was needed to ensure that the correct object was measured among 
many others in these fields. We confirmed a number of identifications with 
IRAC 3.6 \micron\ images from program 35 (Multiplicity and Infrared Colors of Nearby MLT 
Dwarfs, \citet{pat06}). The 24 \micron\ photometric aperture was centered on the photocenter 
of the target. We used the same procedure at 70 \micron\ when the source was 
clearly detected and there were no confusing sources nearby. In cases where 
the 70 \micron\ centroid position disagreed by more than 5\arcsec\  with the 
24 \micron\ position, we forced the 70 \micron\ aperture to be centered at the 
24 \micron\ position.

     The fluxes we report are based on conversion factors of 1.048 $\mu$Jy
/arcsec$^{2}$/(DN/s) and 16.5 mJy/arcsec$^{2}$/U70 at 24 and 70 \micron, 
respectively (U70 = MIPS 70 \micron\ unit). Table \ref{data_table} lists the 
observed fluxes, their formal 1$\sigma$ uncertainties (not including systematic 
calibration uncertainties) and 3$\sigma$ upper limits where no flux could 
be determined.

     For most of the observed stars, the error in the 24 \micron\ 
photometry is dominated by calibration uncertainty, which contributes 
systematic 24 \micron\ errors at the level of $\sim$5\% \citep[see e.g.][]{bei05a}. 
However, we do not depend on the absolute calibration of the measurements 
to identify debris disks. We are instead limited by the scatter in the 
predictions of photospheric flux density from shorter wavelengths \citep[see][]{su06}. 
Background contamination by galactic cirrus or by extragalactic confusion is 
of relatively little importance. Dim sources, however, are limited by photon 
noise and by flat fielding uncertainties created by latent dark images. The 
effective noise level for the dimmest sources is $\sim\!\! 0.05$ mJy. Except
for the faintest sources, we can detect excesses down to $\sim$15\%
above the photosphere in this band.

     At 70 \micron\ the flux error is a combination of calibration uncertainty and 
background noise. The variable sky background, a combination of galactic 
cirrus and extragalactic confusion, presents a noise floor that cannot be 
improved with increases in integration time. This confusion noise is reached 
with integrations of 100's to 1000's of seconds depending on the local level 
of galactic cirrus brightness \citep{bry05}. This variation of the sky background 
within each frame, when convolved with our chosen beam size, is listed as the 
70\micron\ flux error in Table 1. On top of this background noise, we have included 
a 10\% systematic error at 70 \micron, consistent with the uncertainty found within 
photospheric measurements of FGK stars \citep{bei05a,bry05}. A thorough discussion 
of the uncertainties in the 70 \micron\ data can be found in these papers and in 
\citet{gord04}. Excesses of 30\% above the predicted photosphere for bright stars 
at 70 $\mu $m should be readily detectable. 

At 160 \micron, the DAT was used to calibrate the data, and mosaic the images at a 
pixel scale of 8\arcsec/pixel (half the native  scale). The data were flux-calibrated 
using the standard conversion factor of 42.6 MJy/sr/MIPS160, where MIPS160 are the 
instrumental units that result after the data are normalized by the stim flashes. 
The uncertainty on the absolute calibration is 12\% \citep{stan07}. The MIPS 
160 \micron\ array suffers from a near-IR spectral leak that produces a strong ghost 
image overlapping the true 160 \micron\ image for sources with stellar temperatures 
and J magnitudes brighter than 5.5 (MIPS data handbook, http://ssc.spitzer.caltech.edu/mips/dh/).  
Only four of our brightest stars have a possibility of being effected by this leak. 
One of these stars (LHS 70) is the single case where one of our targets was detected 
at 160 \micron, the spectral leak was detected and not the true 160 \micron\ emission. 
This detection is ignored and we give $3\sigma$ upper limits on the 160 \micron\ fluxes 
for all observed targets.

In most cases these limits are estimated from the scatter of pixels falling within a 
sky annulus with radii of 64\arcsec\ -- 128\arcsec, and normalized  to an aperture 
radius of 24\arcsec. In some cases we also estimated the sky noise by dropping multiple 
24\arcsec\ apertures on the images (away from the source position), and using the scatter 
of the fluxes in those apertures instead. The limits are corrected for the finite aperture 
size using a correction factor of 2.37, derived from STiny Tim model PSFs \citep{krist05} 
fitted to observed PSFs \citep{stan07}. The resulting 3$\sigma$ upper limits 
are presented in Table \ref{data_table}. There is considerable variation in the sensitivity 
achieved, due to the effects of galactic cirrus emission. The median upper limit is 74 mJy, 
but the limits for individual stars range from 53 to 963 mJy.

\section{Interpretation}\label{interp}

     Because the intrinsic mid- and far-infrared properties of M stars are not well known, 
we take the approach of using our photometry and existing K band photometry to compare the 
M stars among themselves to look for evidence of excess emission. Extrapolations from K$_S$ 
to [24] require an accurate understanding of the photospheric colors that we determine in 
Section \ref{colorsnxs}. We find no convincing evidence for any 24 \micron\ excesses above 
the photospheric outputs. In section \ref{70xs} we use the ratio of our measured 70 \micron\ 
flux to a 70 \micron\ flux predicted from our 24 \micron\ measurement as our diagnostic tool. 
Again, we find no convincing evidence for a 70 \micron\ excess 
at greater than the 3$\sigma$ level. The 160 \micron\ upper limits are discussed in 
section \ref{160ul}. Finally, in sections \ref{dust} and \ref{comp}, we compare 
the results on dust excesses around M stars with those obtained by previous {\it Spitzer} 
studies of main sequence stars in the solar neighborhood.

\subsection{Photospheric colors and 24 \micron\ excess}\label{colorsnxs}
\subsubsection{Photospheric colors}\label{colors}

     Table \ref{color_table} lists the K$_S$ band magnitudes and the 
colors derived from the infrared photometry. The K$_S$ band data were 
obtained mostly from the 2MASS catalog \citep{cut03a} but in some cases 
K measurements were obtained from other sources and transformed to 2MASS 
K$_S$ (see notes to Table \ref{color_table}). 

\begin{figure}
\epsscale{1.0}
\plotone{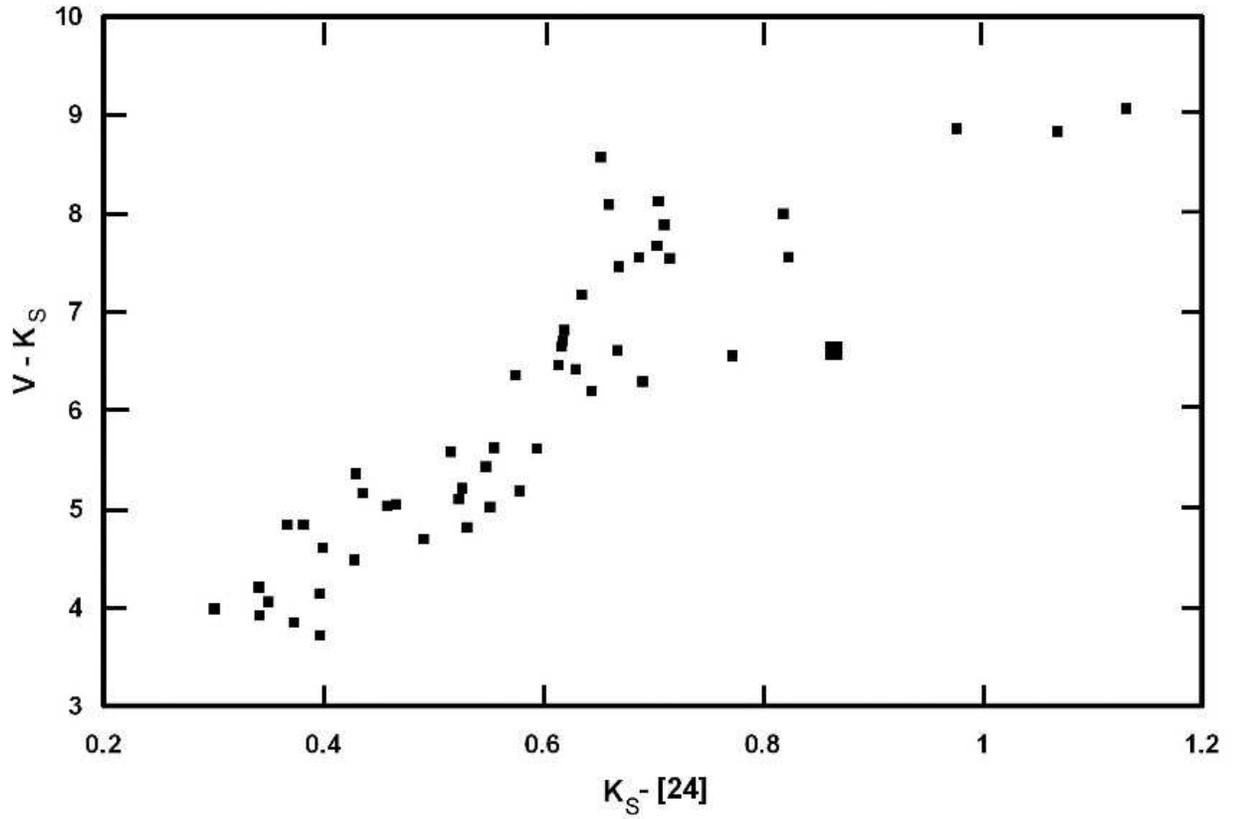}
\caption{
K$_S$-[24] colors vs. V - K$_S$.  The large symbol is for V1581 Cyg B, the only object
with a possible excess at 24$\mu$m identified through this test. 
\label{V-K_fig}}
\end{figure}

     We show a color-color plot of K$_S$ - [24] vs. V - K$_S$ in Figure \ref{V-K_fig}. 
The V magnitudes were obtained from NStars (http://nstars.nau.edu/), where possible, 
with SIMBAD as a secondary 
source. Eleven of our 62 objects have no accurate V photometry available. 
The K$_S$ - [24] colors are greater than zero, reflecting the fact 
that the spectral energy distributions of M stars do not transition to the 
Rayleigh-Jeans regime until well beyond 2 \micron. They also become increasingly 
red with later spectral type, reflected by larger V - K$_S$. With its long
wavelength baseline, the V - K$_S$ color should be monotonic in spectral type/effective
temperature. This plot should therefore allow us to identify excess candidate
stars as outliers toward red K$_S$ - [24] from the overall trend. In fact, there is
one such object, V1581 Cyg B. However, the figure also shows significant
scatter, substantially more than expected purely from the errors in V, 
K$_S$ or [24], suggesting that there are significant variations in V - K$_S$
outside a purely monotonic relationship. 

\subsubsection{Identification of Excesses Using IRFM for Stellar Temperatures}

     Although accurate photometry for the entire sample is available in the near
infrared, the issues shown in Figure 1 cannot be addressed with these measurements
because M dwarf spectral type is not a sufficiently strong monotonic function of infrared colors. 
To improve over color-color plots, we have used a modified form of the infrared flux method \citep[and references 
therein]{blac86} to determine effective temperatures for the stars in the sample, 
designated T$_{ir\kern-0.1em fm}$. We can then plot K$_S$ - [24] as a function of T$_{ir\kern-0.1em fm}$
rather than V - K$_S$. 

The effective temperature can be expressed as:

\[
10log\left( {T_{ir\kern-0.1em fm}}  \right) = 42.32 - 5{\kern 1pt} log\left( {9.43 \times 
10^{15}\left( {\frac{{S_{\nu}  \left( {24} \right)}}{{\left( {\int {S_{\nu}  
\,d\nu} }  \right)^{1/4}\alpha \left( {T} \right)\,\beta \left( {T} 
\right)}}} \right)^{2/3}} \right) - \left( {m_{K} + BC_{K}}  \right),
\]

\noindent
where $\alpha$(T) is a term of order unity that is the ratio of the Planck 
function to the Rayleigh Jeans approximation at 24 \micron, and $\beta$(T) 
is another term of order unity ($\beta$(T) = 0.76 independent of 
temperature for these stars) that is the ratio of the flux received through 
the 24 \micron\ filter to the bolometric flux from the star, divided by the 
similar ratio for a blackbody at the effective temperature of the star. For 
stars of type M6 and later, we take bolometric corrections from \citet{golim04}. 
For stars earlier than M6, we used 

\[
BC_{K} = 0.42 + 1.45\left( {I - K} \right) - 0.20\left( {I - K} 
\right)^{2}.
\]

\noindent
This expression is based on that given by \citet{rh00}, but has been modified 
to join better to the expression of \citet{golim04} at the M5.5 to M6 transition. 
Our derived values of T$_{ir\kern-0.1em fm}$ are listed in Table \ref{color_table}.

     Figure 2 shows the trend of K$_S$ -[24] color vs. 
T$_{ir\kern-0.1em fm}$. The scatter in the observed colors at a fixed temperature 
is small, consistent with it arising purely from the measurement errors. 

     We have compared the colors of the low mass stars with synthetic colors 
generated from a grid of models to be published by Brott et al. \citep[see][]
{brot05}. These models represent dusty atmospheres and are termed the ``GAIA grid" 
because they were calculated in support of planning for the GAIA mission. 
They incorporate advances in the models published by \citet{all01}, through 
improvements in parameters such as opacities and mixing lengths. The theoretical 
spectra were convolved with the wavelength-dependent instrumental response for the 
2MASS K$_S$ filter and the MIPS 24 \micron\ filter. Similar convolutions were 
performed for a Kurucz model A-star spectrum, and the two were ratioed to obtain 
colors relative to A-star colors for the low mass stars. 

\begin{figure}
\epsscale{1.0}
\plotone{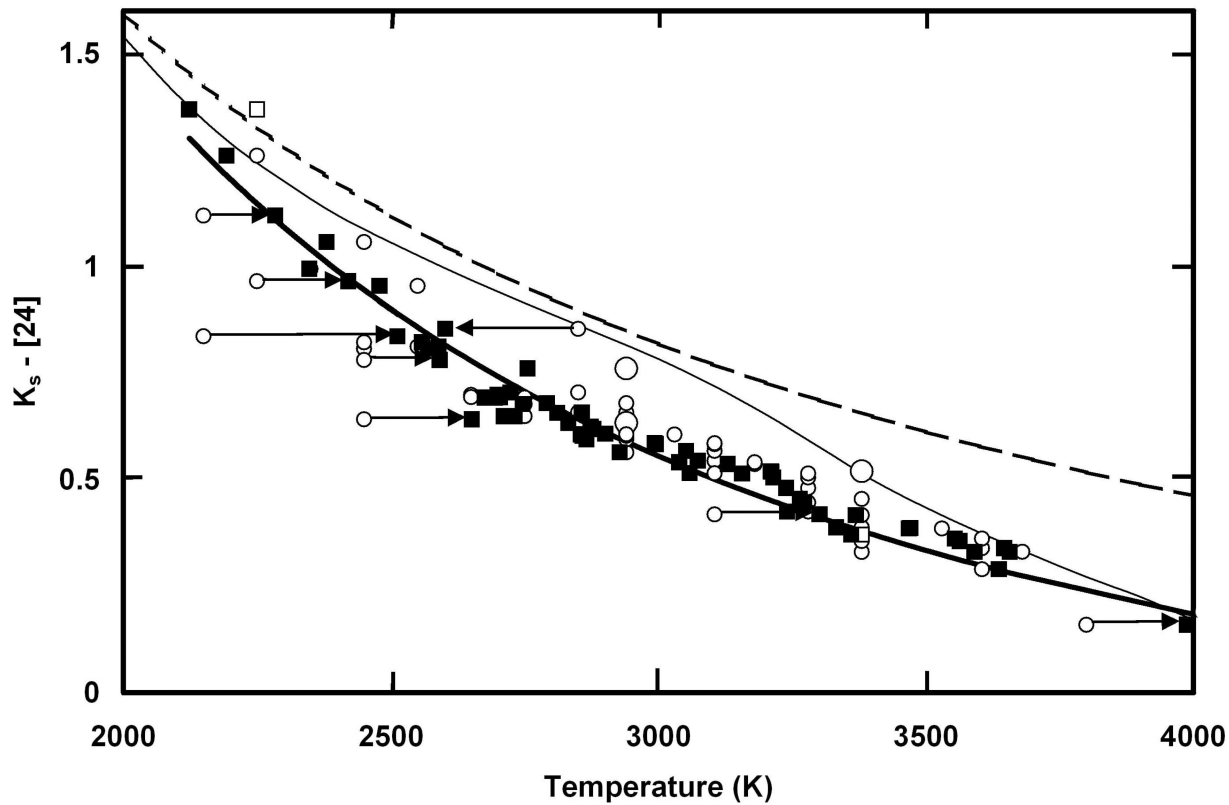}
\caption{
K$_S$-[24] colors vs. T$_{eff}$ and T$_{ir\kern-0.1em fm}$.  The open circles show T$_{eff}$ from the 
spectral type, while the filled boxes are the result of deriving T$_{ir\kern-0.1em fm}$ from the 
infrared flux method, as described in the text. Unresolved binaries are indicated by larger 
circles and young objects by open squares. Cases with large unexplained differences are 
connected with arrows from the spectral estimate to the infrared flux method one.
The thin solid line shows the prediction from the atmospheric models of \citet{brot05}. 
The thick solid line is the fit to K$_S$-[24]vs. T$_{ir\kern-0.1em fm}$ derived in this paper. The 
dashed line shows the color of a black body.
\label{K-24_fig}}
\end{figure}

     The synthetic colors are also compared with the measured ones in
Figure \ref{K-24_fig}. The observations follow roughly the theoretical curve, 
although the observed colors are systematically up to 20\% less red in the 2500 - 3500 K 
range.  Among the contributors to this discrepancy could be the effects 
of dust in the atmospheres, which can modify the absorption feature 
strengths in the K-band as well as affect the radiation transfer and hence the 
emitted flux in the 24 \micron\  band. These new observations should help provide a basis 
for improvement of models.

     Figure \ref{K-24_fig} also plots the positions of the stars using a conventional 
spectroscopic temperature calibration. We took spectral types from the NStars tabulation as of 
December 31, 2006 wherever possible. In the remaining cases, the types are from 
SIMBAD as of the same date. In the latter instances, we consulted the primary references 
to verify the tabulated values. We used a type of M7.5 for LP 044-162, where the SIMBAD 
classification of M6.5 does not agree with the M7.5 found in the literature 
\citep{giz00,cruz03}. From M0 through M6, the temperature calibration is from 
\citet{tok00}, while from M6.5 through M9.5, we have assumed a linear relation 
with a slope of 200K per stellar sub-type. This relation joins the stellar and brown dwarf 
temperatures smoothly and is consistent in approach with the temperature scales for L 
dwarfs \citep{bur02}. Arrows indicate the shifts in effective temperature from 
these values to those determined through the IRFM in this work. If an object has a 
deficiency in flux at 24 \micron\ relative to the conventionally determined temperature, 
then the IRFM calculation will indicate a higher temperature; if the object has a higher 
flux, then the IRFM will return a lower one. It can be seen that most of the shifts are 
toward {\it higher} temperatures. 

If these stars had a combination of underestimated temperature plus an excess at 24 
\micron, then the corrected temperatures should show considerable scatter in the plot 
(because the bolometric corrections on which the plot depends are based on a broad range 
of wavelengths). Instead, these points come to a well determined relationship with only 
small scatter. Therefore, we look for excesses among the stars where the IRFM shifts the 
temperature {\it lower} than the conventional calibration. An example is V1581 Cyg A , 
the star with a spectroscopic temperature of 2940K, adjusted to 2756K by the IRFM. 
However, this star has a low mass companion V 1581 Cyg C that is not resolved in our 
beam. This companion is faint enough in the visible not to bias the spectrum 
significantly, but it can contribute significantly at 24 \micron\ \citep[e.g.,][]{dr00}. 
A satisfactory fit to the properties of V 1581 Cyg A/C is a M6 primary 
and L1 secondary. The predicted brightness difference between A and C is 3 magnitudes at 
R, while the predicted difference at [24] is only 0.8 magnitudes. The observed value is 
3.3 magnitudes at V (red version of V with an effective wavelength of 0.583 \micron; 
\citet{hen99}); a difference of 0.8 magnitudes at [24] explains 
the behavior in the figure. Thus, the behavior in both bands is explained
by the M6/L1 binary hypothesis. Although the excess in this case is not related to a 
debris disk, the temperature shift illustrates how a debris disk could be identified. 

The second object where we derive a lower temperature is V 1581 Cyg B, the single
star suggested to have an excess in the color-color plot in Figure 1. It is not known 
to have a companion, but our results suggest it would be worthwhile to look for one. 
The third strongest example, LHS 3814, is also a double that is unresolved in our beam 
at 24 \micron\ (we have plotted it at the 
temperature of the hotter member). The fourth strongest example, LP 944-20, is thought 
to be only $\sim$300 Myr old, much younger than a typical object in the sample 
\citep{rib03}. Assuming it has lower surface gravity than the other members of the sample, the 
molecular bands in the 2 \micron\ window will be strengthened consistent with its color 
deviation. All of the remaining objects fall in the region where no infrared excess is 
indicated.

\subsubsection{24 \micron\ excesses}\label{24xs}

     Any star with a 24 \micron\ excess should lie above the photospheric color 
locus in Figure 2 or be shifted downward in temperature with our IRFM
estimates.  Although one could hypothesize that V1581 Cyg B has a debris-disk 
excess, since 24 \micron\ excesses are very uncommon around FGK stars \citep{bry05} 
and since we find no large excess at 70 \micron, this possibility is unlikely, and we 
will not consider it further. We therefore interpret Figure \ref{K-24_fig} to show 
that there are no debris-disk-associated excesses at 24 \micron.

     We have repeated our analysis with artificial excesses at 24 \micron. An excess 
of 20\% above the stellar photosphere results in a downward adjustment of the indicated 
stellar temperature by about 200 K. We can compare this value with the scatter in stellar 
colors. The thick solid line in Figure \ref{K-24_fig} can be fitted by

$$ {\rm K}_S-[24]= -0.2915 + {1 \over{1.267\times 10^{-4}\;{\rm T} + 
                   (2.389\times 10^{-4}\;{\rm T})^2 + (2.2\times 10^{-4}\;{\rm T})^3}}$$

\noindent A predicted 24 \micron\ flux density (P$_{24}$ in Table \ref{color_table}) was 
produced from this K$_S$ - [24] and the measured K$_S$. Then the ratio of the measured 
24 \micron\ flux density to P$_{24}$ was formed for each star (F$_{24}$/P$_{24}$ in 
Table \ref{color_table}). This ratio has a mean of 1.04 and a standard deviation of 
0.04 over our sample of 55 stars with derived T$_{ir\kern-0.1em fm}$.  This standard 
deviation corresponds to about 100 K in temperature. Excepting the two cases discussed 
in the preceding section, all the stars in this sample either have spectroscopically 
determined temperatures lower than T$_{ir\kern-0.1em fm}$, or in agreement with this 
temperature. Therefore, the limits on excesses are of the order of 20\%, 2$\sigma$.

\subsection{70 $\mu $m Excesses}\label{70xs}

     Thirteen of the 20 stars with detected 70 \micron\ fluxes have high 
quality measurements (SNR $>4$ and no interference from nearby objects or cirrus) 
that allow a good determination of 70 \micron\ excess. 
We formed the ratio, given in Table \ref{color_table} as F$_{70}$/P$_{70}$, 
of the measured 70 \micron\ flux to a predicted 70 \micron\ flux derived from 
a black body extrapolation of our measured 24 \micron\ flux using T$_{ir\kern-0.1em fm}$ as 
the black body temperature. In Figure 3 we have plotted this 
ratio  vs.~T$_{ir\kern-0.1em fm}$ for these 13 stars.  Objects with only photospheric 
emission appear at a value of one on this plot, while those with excess emission 
at 70 \micron\ would appear at values above unity. None of the 13 stars shows a 
70 \micron\ excess larger than the 3$\sigma$ uncertainty in its (F$_{70}$/P$_{70}$) 
ratio. A 70 \micron\ excess of 100\% of the photospheric emission should be apparent 
in any of these stars. This null result is just at an interesting level of significance 
(see section 3.5.1 for discussion).

\begin{figure}
\epsscale{1.0}
\plotone{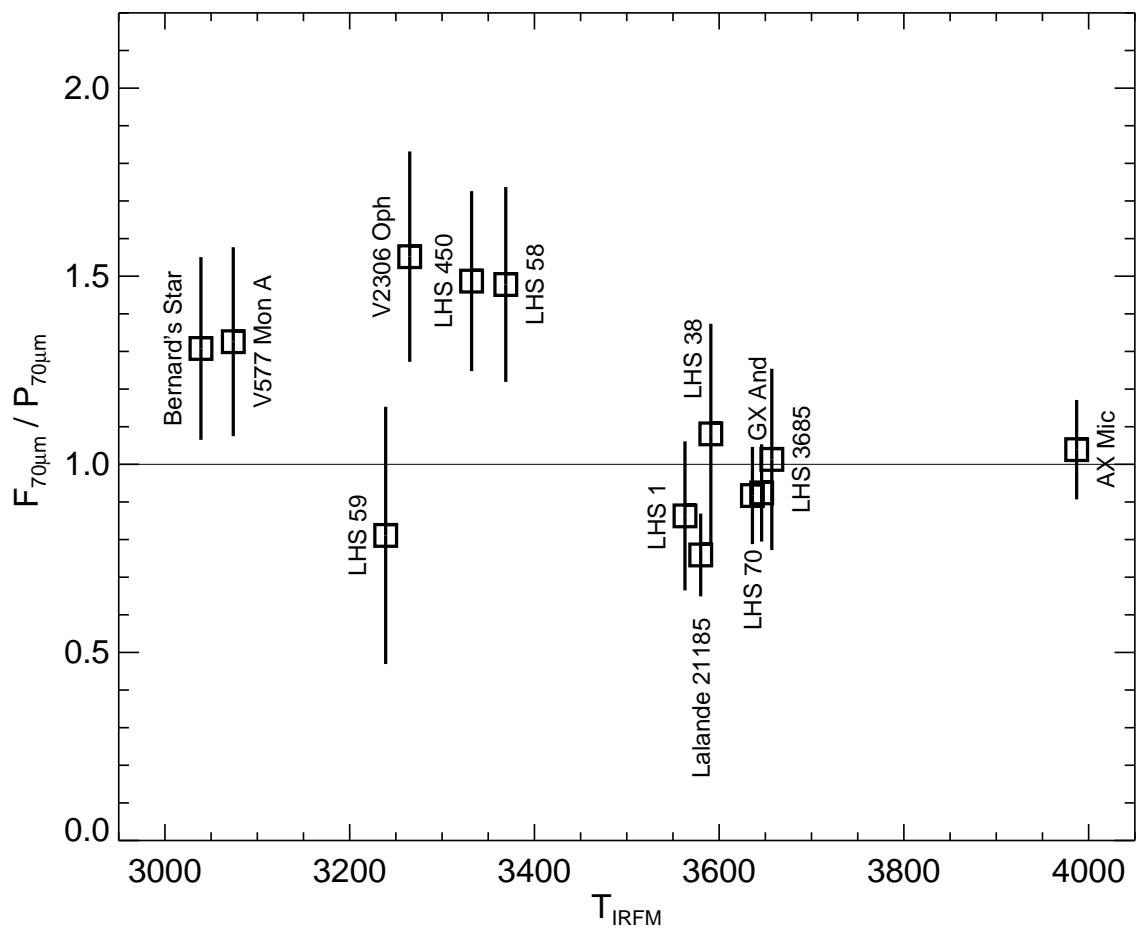}
\caption{
The ratio of measured to predicted 70\micron\ flux vs. the effective temperature of the 
sources determined from the infrared flux method. Only sources that were unambiguously 
detected at 70 \micron\ are included (i.e. SNR $> 3$ and no obvious indications of 
contamination from background sources; see notes to Table \ref{color_table}). 
The 70 \micron\ prediction is extrapolated from the measured 24 \micron\ flux using a 
black body at T$_{ir\kern-0.1em fm}$ for each star, as described in the text.
\label{70_ratio_fig}}
\end{figure}

     We also looked for 70 \micron\ excesses by comparing our 70 \micron\ detections 
among themselves. Any star with an excess should appear out of family with the, 
presumed, majority of non-excess stars. A $\pm 3\sigma$ range for F$_{70}$/P$_{70}$ 
of (0.59, 2.18) was obtained for the stars in Figure \ref{70_ratio_fig} by finding 
the mean and standard deviation of $log({\rm F}_{70}/{\rm P}_{70})$. None of the 13 
stars in Figure \ref{70_ratio_fig} exceed this range, confirming the conclusion of the previous 
paragraph. Among all twenty 70 \micron\ detections (SNR $\ge 3.0$) only AD Leo, YZ Cet and LHS 39 
fall outside of $0.59<{{\rm F}_{70}/{\rm P}_{70}}<2.18$, the range indicating no significant excess. 
AD Leo is anomalously low and may indicate a problem with the measurement of this star. 
YZ Cet may be contaminated by a background galaxy and should not be considered as having an excess 
without confirmation. LHS 39 is tantalizingly high, indicating the possibility of a 70 \micron\ excess, 
but its 70 \micron\ detection is only barely at 3.0 $\sigma$ significance and additional 
uncertainties entering the F$_{70}$/P$_{70}$ ratio result in too low a significance to 
claim an excess.

\subsection{160 $\mu $m Upper Limits}\label{160ul}

     The 160 \micron\ band is more sensitive to the emission from colder dust, 
though the constraints vary greatly from star to star depending on the quality 
of the 160 \micron\ data. Lalande 21185 and V1216 Sgr represent the range of 160 \micron\ 
limits within our data. We discuss the 160 \micron\ upper limits along with
the shorter wavelength measurements in the following section.

\subsection{Fractional dust luminosities}\label{dust}

\begin{figure}
\epsscale{1.0}
\includegraphics[scale=0.6, angle=-90]{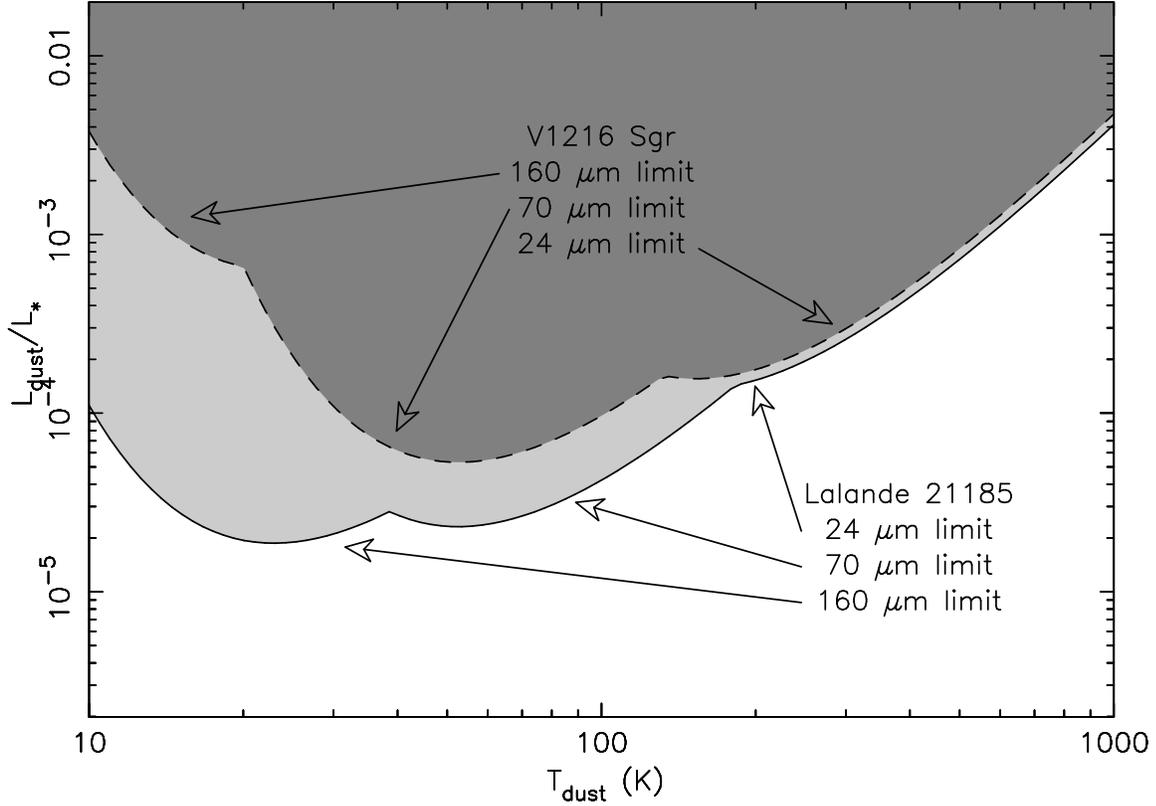}
\caption{
Detection limits for circumstellar dust. The MIPS detection levels are 
shown for two stars as a function of dust luminosity and temperature. 
The light grey region is ruled out by the detection upper limits for 
Lalande 21185 (solid line), a relatively bright star with a clean 
160 \micron\ image, while the dark grey region is ruled out for V1216 Sgr 
(limits shown as a dotted line), a star located in a very noisy 160 \micron\ 
field. Both stars have 1$\sigma$ errors of $\sim $5\% at 24 \micron, but their 
relative accuracies diverge at longer wavelengths; 1$\sigma$ errors at 
160 \micron\ are 0.9 and 70 times the stellar photosphere for Lalande 21185 
and V1216 Sgr respectively. In both cases, 3 $\sigma $ limits are plotted. 
\label{dust_fig}}
\end{figure}

     Figure \ref{dust_fig} illustrates the limits on dust luminosity derived from our flux 
measurements. The upper limits on the ratio of dust luminosity to star luminosity 
(L$_{d}$/L$_{*}$) are shown as a function of dust temperature for two stars, 
Lalande 21185 and V1216 Sgr, representing the range of limits obtained for stars 
detected at 70 \micron. For each dust temperature, we calculate the maximum 
blackbody emission that is consistent with the observed 3$\sigma$ limits. 
Each of the curved segments in Figure \ref{dust_fig} corresponds to an individual observation 
at a specific wavelength, with the temperatures that an observation is most sensitive 
to depending on that wavelength. The 70 \micron\ observations are particularly 
useful for constraining the emission from dust with temperatures $\sim$50 - 100 K, 
for which fractional luminosities greater than $\sim\!\! 10^{-4}$ are generally excluded.

     The constraints on L$_{d}$/L$_{*}$ in Figure \ref{dust_fig} can be translated into upper 
limits on dust mass, for an assumed dust size and orbital location. Assuming 
micron-size dust, the 160 \micron\ upper limits correspond to dust mass 
limits of $\sim\!\! 10^{-6}$ to $10^{-4}$ M$_{Earth}$ of 20 K dust at 50 AU. 
Observations with moderate S/N at 70 \micron\ translate to a maximum of 
$\sim\!\! 10^{-7}$ M$_{Earth}$ of 50 K dust at 5 AU, while 24 \micron\ observations 
limit warm, 250 K dust to $\sim\!\! 10^{-9}$ M$_{Earth}$ at 0.2 AU (again 
assuming micron-size dust in both cases).

     While fractional dust luminosities similar to that of $\beta$ Pic 
(L$_{d}$/L$_{*} = 3\times 10^{-3}$) would have been detected in all stars with 
their photospheres detected at 70 \micron, luminosities as low as that of Vega 
(L$_{d}$/L$_{*}$$\sim 2\times 10^{-5}$) could only be detected around stars 
with optimal data such as Lalande 21185.

\subsection{Comparison of M stars with F, G, and K Stars}\label{comp}

\subsubsection{Previous Work}\label{prev_work}

     Debris disk excesses in M stars have been elusive, partly because of the 
increase in detection difficulty with decreasing stellar luminosity (see 
Section \ref{quant_comp}). \citet{song02} and \citet{plav05} have searched 
for excesses from M stars from IRAS data complemented by ground based 
measurements. Both studies find no credible excesses around mature M stars, 
that is those not associated with known young clusters nor showing spectroscopic signs 
of youth. Our results confirm this conclusion, using the more sensitive and accurate 
{\it Spitzer} data. Using reprocessed IRAS data, \citet{back86}, 
reported an infrared excess from Ross 128 at 60 and 100 \micron. This was 
not confirmed by subsequent processing in the IRAS Faint Source Survey, and 
is also contradicted by the {\it Spitzer}/MIPS results. The 0.12 Jy excess they 
reported at 60 \micron\ is 8 times larger than our 3 sigma upper limit at 
70~\micron, and is therefore confidently excluded by our data. Using {\it Spitzer} 
spectroscopy, \citet{jur04} report no excess for Ross 128. This leaves no 
known example of a mature main sequence M dwarf known to possess a 
far-infrared excess.

     It is interesting to compare the present results with similar studies of 
stars of other spectral types to see whether M star debris disks 
are less common than disks around earlier type stars. As a comparison 
sample, we have used both the study of F5-K5 main-sequence field stars 
discussed by \citet{bry05} and the similar set of stars studied by 
\citet{kim05}. We combine all the data without regard to whether they apply
to single stars, binaries, or stars with radial velocity planets. 
There are 69 stars in Bryden et al. and 35 in Kim et al., 
for a total of 104. Of these stars, 12 have detected excesses. We need
to add $\sim$2 more detections to allow for the weighted contribution of the
four stars with bright excesses detected by IRAS and not re-observed 
\citep[see][]{bry05}. We find that $13 \pm 4$\% of these stars have excesses 
greater than 100\%. Only one of the 69 stars in \citet{bry05}, HD 69830, 
shows evidence for a significant 24 \micron\ photometric excess more than 15\% 
above the photosphere. \citet{kim05} included only stars with cold excesses, 
so we cannot derive an independent statistic in this regard. However, it is 
clear that 24 \micron\ excesses are uncommon in solar type main sequence field stars. 
We find no convincing 24\micron\ excess in a sample of 56 M stars. 

     The current sample of 20 M dwarfs with $3\sigma$ or better 70 \micron\ detections is quite 
small for purposes of drawing confident statistical inferences of debris disk 
frequency. Binomial statistics indicate that our result of no disk detections 
in the 20 stars only allows us to derive an upper limit of 14\%, with 2$\sigma$ 
confidence, to the incidence of disks among M dwarfs within 5pc of the Sun. This upper limit
is right at the measured incidence level for more massive stars. 
Our upper limit is also consistent with a value of 13\% reported by \citet{les06}\ for 
20 to 200 Myr old field M dwarfs. However, we note that our 5 pc sample 
is likely to contain a large fraction of stars older than 200 Myr and may not be directly 
comparable.

     \citet{song02} find that a number of extremely young M stars have 
detectable excesses (i.e., Hen 3-600 and AU Mic, both less than 15 M yr 
old). \citet{low05} report sensitive {\it Spitzer} measurements in the young 
TW Hya association where, of the 24 stars observed, 5 have significant 
excess at 24 \micron\ and 6 have excess at 70 \micron. \citet{chen05} confirm 
the excess in AU Mic. Thus, excesses are not uncommon in late-type stars, 
including M-type ones, at an age of $\sim$10 Myr. However, they virtually 
completely disappear in mature low-mass stars.

\subsubsection{Approach for Quantitative Comparison}\label{quant_comp}

     To expand on the results in the preceding section, we need to use an appropriate 
metric to characterize debris systems. We will adopt the 70 \micron\ fractional 
excess, F$_{excess}$/F$_{photosphere}$, where F$_{excess}$ is the flux 
density of the infrared excess and F$_{photosphere}$ that of the star itself, 
both at 70 \micron. We use the 70 \micron\ fractional excess as a proxy for the 
fractional luminosity, L$_{IR}$/L$_{star}$, because it should be representative of 
disk luminosity for objects with no excess at shorter wavelengths. 

     As stars become less luminous and of lower temperature, it becomes 
increasingly difficult to measure infrared excesses to a given limit in 
fractional luminosity. This issue can be understood by considering a 
set of stars with varying temperature. The total luminosity of these objects, 
and thus the dust luminosity, will vary as ${\rm T}^4 {\rm R}^2$ (where R is 
the stellar radius), but their photospheric outputs in the far infrared will be 
in the Rayleigh-Jeans regime and will only go as ${\rm T} {\rm R}^2$. Thus, for 
a given threshold for excess detection above the photosphere, the fractional 
luminosity threshold will go as T$^{-3}$.

\subsubsection{Application to M Star Sample}\label{Mstars}

     To obtain a quantitative measure of the average level of excess 
in the M star sample, we proceed in a manner analogous to stacking, a 
technique often used to enhance the average signal to noise ratio with 
observations of a suite of similar sources. Our various M stars 
have different expected signal levels and, as a result, the measurements 
are of intrinsically differing ratios of signal to noise. 
Therefore, we stack by taking the weighted 
average of the measured excess ratios at 70\micron. We eliminate from 
this average all objects with a nominal SNR $<$ 4. We adopt for the 
remaining sources a net error consisting of the statistical error 
indicated in Table 1, plus a systematic error of 6\% to reflect the 
scatter in measurements of bright objects at 70 $\mu $m with MIPS 
\citep[e.g.][]{gord05}. The result is a weighted average excess ratio 
at 70 \micron\ of $1.02 \pm 0.05$; that is, on average, there is no excess 
from these stars. We place a $2\sigma$ upper limit of 1.12 to the 
average excess ratio, or 0.12 to the excess (relative to the photospheric 
emission). We also compute an average effective temperature for this sample 
as a weighted average of the individual stellar effective temperatures 
where the weights are those used for the excess ratios. The result is 
T = 3555 K. 

     As a comparison sample, we use the FGK stars in \citet{bry05}. We add 
in the four additional stars listed by Bryden et al. but measured in 
other {\it Spitzer} programs at half weight, since Bryden et al. are 
reporting on about half of their sample. We also add in DY Eri and eta Cas 
from this program. The result is an average excess ratio of $1.37 \pm 0.033$. 
The average temperature of this sample computed in the same way 
as for the M stars is 5763 K. If we scale the average excess for this sample 
as T$^3$, we predict that the average for the M stars if they have the 
same fractional luminosity in infrared excess emission is 0.09, just within 
the upper limit from our observations of 0.12. 

     Another effect that might lead to more prominent debris excesses 
in M stars is the scaling of dust mass to produce a given excess level, 
as a function of stellar luminosity. For a given fractional luminosity upper 
limit at a given wavelength, the minimum mass limits derived from the M star 
observations reported here are lower than would be the corresponding limits from 
more luminous stars. The reason is that the fractional dust luminosity is 
proportional to the total solid angle subtended by the dust as viewed from 
the star and that the temperature of the dust increases with the star's 
luminosity but decreases with distance from the star. Thus, at a fixed dust 
temperature, an increase in stellar luminosity requires that the dust be more 
distant which, in turn, requires more dust to subtend enough solid angle to 
keep L$_{d}$/L$_{*}$ constant. In the simple case of a single fixed grain 
size, the number of grains to achieve a certain fractional luminosity at a 
given temperature scales with the luminosity of the star. Thus the minimum 
dust mass associated with an M star with $\sim 0.01\;{\rm L}_\odot$ solar luminosities 
would be $\sim 1$\% of the corresponding minimum mass associated with a 
solar mass star, again assuming the same limit to the fractional luminosity 
at a given wavelength. For typical M stars in the present sample, therefore, 
it is not surprising to note that the limits on 24 \micron\ excesses correspond to 
masses as low as 10$^{-9}$ earth masses, while at 70 \micron, the non-detections 
could imply masses below 10$^{-7}$ earth masses. Similar {\it Spitzer}/MIPS 
surveys for IR excess around FGK stars \citep[e.g.][]{bry05} can detect 
dust with lower luminosity relative to the central star 
(L$_{dust}$/L$_{star}$ as low as $5 \times 10^{-6}$), but are less sensitive in 
terms of overall dust mass. The corresponding limit for 
non-detections of 70 \micron\ excess around FGK stars is $>10^{-6}$ M$_{Earth}$, an 
order of magnitude above that for M stars. 

\subsubsection{Discussion}\label{complex}

     The fractional luminosity is the 
appropriate metric if we assume that hypothetical debris disks would have 
the same covering fraction independent of stellar type, thus absorbing and 
reradiating in the far infrared an equivalent fraction of the stellar output. 
Although the fractional luminosity provides a reasonable first-order metric 
to compare excesses in different stellar populations, there are many other 
considerations. For example, one would expect that cool, low-mass stars 
might be less effective at heating circumstellar dust than hotter ones, even 
after allowing for the difference in luminosity. The degree to which this 
expectation is realized depends on the optical properties of the debris disk 
particles. They are likely to be relatively large grain aggregates, whose 
behavior may differ substantially from that of typical interstellar grains. 
The grain loss mechanisms - photon pressure, Poyting-Robertson drag,
and corpuscular winds - are generally different in low mass stars than 
in high mass ones, and may cause significant differences in their disks
(e.g., strong winds may destroy them quickly). Also, our study 
probes debris disks at 70 \micron\ and hence at a 
temperature of about 50 K. The equilibrium distance from a star of luminosity 
L for material at a given temperature scales as the square root of L. Given 
the steep dependence of stellar luminosity on mass, we are probing 
distinctly different physical regions around the M stars in this sample than 
the regions probed in observations of solar-like stars. 

\section{Conclusions}\label{conclude}

     We observed a sample of 62 M stars to search for debris disk dust and to 
study photospheric models. We demonstrate a modified infrared flux method to 
determine the effective temperatures of these stars, taking advantage of the 
high accuracy measurements available at 24 \micron\ with {\it Spitzer}. Using these 
effective temperatures, the K$_S$ - [24] colors of the stars agree 
reasonably well with the theoretical models of Brott et al. (private 
communication). None of the stars in our sample has a high-weight detection 
of a mid- or far-infrared excess. In particular, the sample lacks the 
occasional stars with large excess that typifies samples of more massive 
stars. We compute a weighted average of the observed excess ratios at 70 \micron\ 
as a metric to characterize the excesses in this and other samples 
of stars. We find that the average excess in the M stars is at least four 
times less than the average in a sample of FGK stars. This limit is just at 
the level for equal fractional infrared luminosity in the two samples. 
We have set low limits ($10^{-4}$ to $10^{-9}$ M$_{Earth}$ depending on location) 
for the maximum mass of dust possible around our stars.

\acknowledgements

     This work is based, in part, on observations made with the {\it Spitzer} Space 
Telescope, which is operated by the Jet Propulsion Laboratory, California 
Institute of Technology under contract with NASA. Support for this work 
was provided by NASA through contract 1255094. 
This work also makes use of data products from the Two 
Micron All Sky Survey, which is a joint project of the University of 
Massachusetts and the Infrared Processing and Analysis Center/California 
Institute of Technology, funded by the National Aeronautics and Space 
Administration and the National Science Foundation. We would further like to 
thank Peter Hauschildt for supplying us with results from the latest GAIA model
spectra of late type stars.

\clearpage

\begin{deluxetable}{llrllccc}
\tabletypesize{\scriptsize}
\tablecaption{Observational Data \label{data_table}}
\tablewidth{0pt}
\tablehead{
\colhead{Name} & \colhead{Gliese}& \colhead{LHS}& \colhead{SpT\tnm{1}}& \colhead{F$_{24}$(mJy)}& 
\colhead{F$_{70}$(mJy)\tnm{*}}& \colhead{F$_{160}$(mJy,3$\sigma $)}
}
\startdata
AX Mic                  & GJ 825    & 66    & M0    & 504.00 $\pm$ 0.57       & 61.8 $\pm$ 3.8       & $<$75    \\
LHS 38                  & GJ 412 A  & 38    & M1    & 123.94 $\pm$ 0.21       & 15.8 $\pm$ 3.9\tnm{a}& $<$67    \\
GX And                  & GJ 15 A   & 3     & M1.5  & 249.45 $\pm$ 0.29       & 27.2 $\pm$ 2.2       & $<$102    \\
Kapteyn's Star          & GJ 191    & 29    & M1.5  &  98.46 $\pm$ 0.14       &  4.6 $\pm$ 2.4\tnm{b}& \nodata  \\
LHS 70                  & GJ 887    & 70    & M1.5  & 436.96 $\pm$ 0.43       & 47.3 $\pm$ 3.9       & $<$53    \\
Lalande 21185           & GJ 411    & 37    & M2    & 481.55 $\pm$ 0.40       & 43.0 $\pm$ 3.5       & $<$62    \\
LHS 1                   & GJ 1      & 1     & M3    & 159.13 $\pm$ 0.28       & 16.2 $\pm$ 3.2       & \nodata  \\
AD Leo                  & GJ 388    & 5167  & M3    & 151.62 $\pm$ 0.20       &  8.4 $\pm$ 2.3       & \nodata  \\
V2306 Oph               & GJ 628    & 419   & M3    & 104.83 $\pm$ 0.23       & 19.2 $\pm$ 2.8       & $<$154   \\
LHS 449                 & GJ 674    & 449   & M3    & 120.51 $\pm$ 0.22       & 10.4 $\pm$ 7.3\tnm{c}& $<$963   \\
LHS 450                 & GJ 687    & 450   & M3    & 160.17 $\pm$ 0.21       & 28.1 $\pm$ 3.4       & $<$83    \\
LHS 58                  & GJ 725 A  & 58    & M3    & 183    $\pm$ 3          & 31.9 $\pm$ 3.3\tnm{d}& $<$73\tnm{d}\\
LHS 3685                & GJ 832    & 3685  & M3    & 163.15 $\pm$ 0.18       & 19.5 $\pm$ 4.1       & $<$57    \\
LHS 3814                & GJ 860 A  & 3814  & M3    & 146.0  $\pm$ 0.4        & 117  $\pm$  61\tnm{e}& \nodata  \\
GQ And                  & GJ 15 B   &       & M3.5  &  49.46 $\pm$ 0.20       &  5.3 $\pm$ 1.9\tnm{f}& \nodata  \\
Luyten's Star           & GJ 273    & 33    & M3.5  & 127.16 $\pm$ 0.22       & 11.4 $\pm$ 3.7       & $<$71    \\
LHS 59                  & GJ 725 B  & 59    & M3.5  & 115    $\pm$ 3          & 11.0 $\pm$ 3.3\tnm{d}& \nodata\tnm{d}\\
V1216 Sgr               & GJ 729    & 3414  & M3.5  &  83.67 $\pm$ 0.27       & 17.0 $\pm$ 5.7\tnm{g}& $<$318   \\
IL Aqr                  & GJ 876 A  & 530   & M3.5  & 108.27 $\pm$ 0.16       &  9.2 $\pm$ 2.2       & $<$67    \\
Ross 128                & GJ 447    & 315   & M4    &  66.32 $\pm$ 0.24       &  7.5 $\pm$ 5.0       & $<$61    \\
Barnard's Star          & GJ 699    & 57    & M4    & 188.45 $\pm$ 0.25       & 29.1 $\pm$ 4.4       & $<$141    \\
YZ Cet                  & GJ 54.1   & 138   & M4.5  &  33.69 $\pm$ 0.19       &  9.0 $\pm$ 2.0\tnm{h}& \nodata  \\
TZ Ari                  & GJ 83.1   & 11    & M4.5  &  27.71 $\pm$ 0.21       &               \tnm{i}& \nodata  \\
DY Eri C                & GJ 166 C  & 25    & M4.5  &  49.05 $\pm$ 0.14       &  4.0 $\pm$ 2.0       & \nodata  \\
V577 Mon A              & GJ 234 A  & 1849  & M4.5  &  77.97 $\pm$ 0.20       & 12.2 $\pm$ 1.9\tnm{j}& $<$128    \\
LHS 451                 & GJ 682    & 451   & M4.5  &  62.15 $\pm$ 0.23       &               \tnm{k}& \nodata  \\
EZ Aqr                  & GJ 866 A  & 68    & M5    &  78.73 $\pm$ 0.19       & -4.2 $\pm$ 3.5       & \nodata  \\
LHS 39                  & GJ 412 B  & 39    & M5.5  &   9.89 $\pm$ 0.16       & 10.1 $\pm$ 3.4\tnm{l}& \nodata  \\
FL Vir                  & GJ 473A   & 333   & M5.5  &  49.4  $\pm$ 0.4        &  3.0 $\pm$ 2.0       & \nodata  \\
Proxima Cen             & GJ 551    & 49    & M5.5  & 235.06 $\pm$ 0.37       &               \tnm{m}& \nodata  \\
Ross 248                & GJ 905    & 549   & M5.5  &  52.78 $\pm$ 0.16       & 10.7 $\pm$ 2.0\tnm{n}& \nodata  \\
LHS 2                   & GJ 1002   & 2     & M5.5  &  14.60 $\pm$ 0.18       &  4.1 $\pm$ 2.6       & \nodata  \\
LHS 1565                & GJ 1061   & 1565  & M5.5  &  29.65 $\pm$ 0.22       &               \tnm{p}& $<$63    \\
V1581 Cyg A             & GJ 1245 A & 3494  & M5.5  &  27.0  $\pm$ 1.6        & -12  $\pm$  10\tnm{q}& $<$387   \\
LHS 1375                & GJ 3146   & 1375  & M5.5 s& 3.31 $\pm $ 0.13        & \nodata              & \nodata  \\
LHS 288                 & GJ 3618   & 288   & M5.5  &  10.71 $\pm$ 0.14       &               \tnm{r}& \nodata  \\
UV/BL Cet               & GJ 65 B   & 10    & M6    &  95.74 $\pm$ 0.22\tnm{r}&  5.6 $\pm$ 1.9\tnm{s}& $<$68    \\
GL 316.1                & GL 316.1  & 2034  & M6   s& 1.35 $\pm $ 0.11        & \nodata              & \nodata  \\
Wolf 359                & GJ 406    & 36    & M6    &  49.90 $\pm$ 0.30       &               \tnm{t}& \nodata  \\
V1581 Cyg B             & GJ 1245 B & 3495  & M6    &  18.0  $\pm$ 0.7        & -12  $\pm$  10\tnm{q}& \nodata  \\
GJ 283B                 & GJ 283 B  & 234   & M6.5 s& 1.05 $\pm $ 0.13        & \nodata              & \nodata  \\
DX Cnc                  & GJ 1111   & 248   & M6.5  &  17.18 $\pm$ 0.22       &  6.0 $\pm$ 3.0\tnm{u}& $<$101    \\
LHS 292                 & GJ 3622   & 292   & M6.5  &   9.41 $\pm$ 0.13       &  1.1 $\pm$ 1.9\tnm{v}& \nodata  \\
LHS 2930                & GJ 3855   & 2930  & M6.5 s& 1.63 $\pm $ 0.1         & \nodata              & \nodata  \\
GJ 644C                 & GJ 644 C  & 429   & M7    & 4.17 $\pm $ 0.12        & \nodata              & \nodata  \\
LHS 3003                & GJ 3877   & 3003  & M7    & 3.75 $\pm $ 0.11        & \nodata              & \nodata  \\
2MASS J09522188-1924319 &           &       & M7   s& 0.53 $\pm $ 0.07        & \nodata              & \nodata  \\
2MASS J23062928-0502285 &           &       & M7.5 s& 1.35 $\pm $ 0.12        & \nodata              & \nodata  \\
LP 044-162              &           &       & M7.5\tnm{2}  & 1.08 $\pm $ 0.08 & \nodata              & \nodata  \\
LP 349-025              &           &       & M8   s& 2.31 $\pm $ 0.12        & \nodata              & \nodata  \\
LP 326-021              &           &       & M8   s& 0.89 $\pm $ 0.09        & \nodata              & \nodata  \\
LP 412-031              &           &       & M8   s& 0.74 $\pm $ 0.07        & \nodata              & \nodata  \\
LP 771-021              &           &       & M8   s& 0.41 $\pm $ 0.05        & \nodata              & \nodata  \\
LHS 2397 A              & GJ 3655   & 2397A & M8   s& 0.99 $\pm $ 0.06        & \nodata              & \nodata  \\
ESO 207-61              &           &       & M8   s& 0.16 $\pm $ 0.04        & \nodata              & \nodata  \\
TVLM 513-46546          &           &       & M8.5 s& 1.03 $\pm $ 0.1         & \nodata              & \nodata  \\
LP 944-20               & \nodata   &       & M9 s  &   3.95 $\pm$ 0.05       &  7   $\pm$   3\tnm{w}& \nodata  \\
LHS 2065                & GJ 3517   & 2065  & M9    & 1.9  $\pm $ 0.09        & \nodata              & \nodata  \\
TVLM 868-110639         &           &       & M9   s& 0.68 $\pm $ 0.03        & \nodata              & \nodata  \\
BRI 0021-0214           &           &       & M9.5 s& 1.26 $\pm $ 0.05        & \nodata              & \nodata  \\
2MASSW J1733189+463359  &           &       & M9.5 s& 0.28 $\pm $ 0.03        & \nodata              & \nodata  \\
DENIS-P J0021.0-4244    &           &       & M9.5 s& 0.26 $\pm $ 0.04        & \nodata              & \nodata  \\
\enddata

\tnt{*}{Entries with an ellipsis (\nodata) in the 70 \micron\ column are from {\it Spitzer} program 56.}
\tnt{1}{Spectral types are from the NStars data base except for those marked ``s'' which are from Simbad.}
\tnt{2}{See text for LP 044-162 spectral type source.}
\tnt{a}{{L}HS 38: Some cirrus at 70 \micron}
\tnt{b}{Kapteyn's: No source apparent at 70 \micron. Photometry at reference position.}
\tnt{c}{{L}HS 449: Bad cirrus confusion at 70 \micron}
\tnt{d}{{L}HS 58 \& 59: 14\arcsec\  separation. 70 \micron\ fluxes from deblending. 160 \micron\ upper limit is for the combination.}
\tnt{e}{{L}HS 3814: Bad cirrus, especially at 70 \micron}
\tnt{f}{GQ And: 35\arcsec\  separation from GX And: possible slight blending at 70 \micron.}
\tnt{g}{V1216 Sgr: Bad Cirrus at 70 \micron}
\tnt{h}{YZ Cet: Diffuse source, probable background contamination}
\tnt{i}{TZ Ari: Bright background source 20\arcsec\  north accounts for flux}
\tnt{j}{V577 Mon A: 2\arcsec\  separation from GJ23B. Photometry includes both sources in both bands}
\tnt{k}{{L}HS 451: Bad cirrus confusion: no source apparent at 70 \micron}
\tnt{l}{{L}HS 39: 70 \micron\ photometry done at 24 \micron\ position. 32\arcsec\  separation from LHS 38}
\tnt{m}{Proxima Cen: Very bad cirrus at 70 \micron.}
\tnt{n}{Ross 248: 70 \micron\ source offset slightly from 24 \micron\ source; possible contamination.}
\tnt{p}{{L}HS 1565: Bright background galaxy. No 70 \micron\ photometry possible}
\tnt{q}{V1581 Cyg A \& B: 8\arcsec\ separation and bad cirrus. 70 \micron\ photometry includes both sources}
\tnt{r}{{L}HS 288: Not on the 70 \micron\ image (error in requested pointing)}
\tnt{s}{UV Cet/BL Cet: 2\arcsec\  separation. Photometry is for both sources in both bands}
\tnt{t}{Wolf 359: Not on the 70 \micron\ image (error in requested pointing)} 
\tnt{u}{DX Cnc: No source apparent at 70 \micron. Photometry at reference position.}
\tnt{v}{{L}HS 292: 70 um photometry at 24 \micron\ position. Nearby 70 \micron\ source.}
\tnt{w}{{L}P 944-20: Faint source at 70 \micron; another 70 \micron\ source 50\arcsec\  north}
\end{deluxetable}
\clearpage

\begin{deluxetable}{lccccccccccc}
\tabletypesize{\scriptsize}
\rotate
\tablecaption{Colors and Derived Quantities \label{color_table}}
\tablewidth{0pt}
\tablehead{
\colhead{Name} & \colhead{Dist(pc)}& \colhead{SpT}& \colhead{T$_{eff}$}& \colhead{T$_{ir\kern-0.1em fm}$}& 
\colhead{B.C.}& \colhead{K$_S$\tnm{*}} & \colhead{K$_S$-[24]\tnm{\dag}} & \colhead{P$_{24}$\tnm{\ddag}} & 
\colhead{F$_{24}$/P$_{24}\tnm{\ddag}$} & \colhead{P$_{70}$\tnm{\ddag}} & \colhead{F$_{70}$/P$_{70}$\tnm{\ddag}}
}
\startdata

AX Mic             & 3.95   & M0   & 3800 &  3987 & 2.370 &  $3.07\pm 0.02$\tnm{a}& $0.19\pm 0.06$ & 502.04 &  1.00 &  58.37 &   1.06   \\
LHS 38             & 4.85   & M1   & 3680 &  3591 & 2.530 &  $4.77\pm 0.02$       & $0.37\pm 0.06$ & 116.58 &  1.06 &  14.44 &   1.09   \\
GX And             & 3.56   & M1.5 & 3605 &  3646 & 2.470 &  $4.02\pm 0.02$       & $0.38\pm 0.06$ & 228.78 &  1.09 &  29.03 &   0.94   \\
Kapteyn's Star     & 3.92   & M1.5 & 3605 &  3554 & 2.474 &  $5.05\pm 0.02$       & $0.40\pm 0.06$ &  88.49 &  1.11\tnm{d} &  11.46 & $< 0.6$\tnm{d}  \\
LHS 70             & 3.29   & M1.5 & 3605 &  3636 & 2.530 &  $3.36\pm 0.02$\tnm{a}& $0.33\pm 0.06$ & 421.41 &  1.04 &  50.86 &   0.93   \\
Lalande 21185      & 2.54   & M2   & 3530 &  3469 & 2.530 &  $3.35\pm 0.02$\tnm{b}& $0.42\pm 0.06$ & 432.61 &  1.11\tnm{d} &  56.10 &   0.77\tnm{d}   \\
LHS 1              & 4.36   & M3   & 3380 &  3563 & 2.530 &  $4.52\pm 0.2 $       & $0.39\pm 0.21$ & 148.05 &  1.07 &  18.55 &   0.87   \\
AD Leo             & 4.89   & M3   & 3380 &  3360 & 2.648 &  $4.59\pm 0.02$       & $0.41\pm 0.06$ & 148.76 &  1.02 &  17.73 &   0.47   \\
V2306 Oph          & 4.24   & M3   & 3380 &  3265 & 2.710 &  $5.08\pm 0.02$       & $0.49\pm 0.06$ &  99.15 &  1.06 &  12.28 &   1.56   \\
LHS 449            & 4.54   & M3   & 3380 &  3471 & 2.580 &  $4.86\pm 0.02$       & $0.42\pm 0.06$ & 112.58 &  1.07 &  14.07 & $< 1.6$  \\
LHS 450            & 4.54   & M3   & 3380 &  3332 & 2.710 &  $4.55\pm 0.02$       & $0.43\pm 0.06$ & 155.96 &  1.03 &  18.74 &   1.50   \\
LHS 58             & 3.53   & M3   & 3380 &  3369 & 2.650 &  $4.43\pm 0.02$       & $0.45\pm 0.06$ & 171.81 &  1.07 &  21.40 &   1.49   \\
LHS 3685           & 4.93   & M3   & 3380 &  3657 & 2.470 &  $4.47\pm 0.02$\tnm{c}& $0.37\pm 0.06$ & 150.66 &  1.08 &  18.98 &   1.03   \\
LHS 3814           & 4.03   & M3   & 3380 &  3211 & 2.648 &  $4.78\pm 0.03$       & $0.56\pm 0.06$ & 132.40 &  1.10 &  17.13 & $<10.7$  \\
GQ And             & 3.56   & M3.5 & 3280 &  3157 & 2.708 &  $5.95\pm 0.02$       & $0.55\pm 0.06$ &  46.12 &  1.07 &   5.81 & $< 1.0$  \\
Luyten's Star      & 3.79   & M3.5 & 3280 &  3273 & 2.710 &  $4.86\pm 0.02$       & $0.49\pm 0.06$ & 119.93 &  1.06 &  14.90 &   0.77   \\
LHS 59             & 3.53   & M3.5 & 3280 &  3239 & 2.710 &  $5.00\pm 0.02$       & $0.52\pm 0.06$ & 106.87 &  1.08 &  13.48 &   0.82   \\
V1216 Sgr          & 2.97   & M3.5 & 3280 &  3215 & 2.710 &  $5.37\pm 0.02$       & $0.54\pm 0.06$ &  76.76 &  1.09 &   9.81 & $< 1.8$  \\
IL Aqr             & 4.70   & M3.5 & 3280 &  3240 & 2.760 &  $5.01\pm 0.02$       & $0.46\pm 0.06$ & 105.85 &  1.02 &  12.69 &   0.72   \\
Ross 128           & 3.35   & M4   & 3180 &  3130 & 2.760 &  $5.65\pm 0.02$       & $0.57\pm 0.06$ &  61.53 &  1.08 &   7.79 & $< 1.9$  \\
Barnard's Star     & 1.83   & M4   & 3180 &  3039 & 2.850 &  $4.52\pm 0.02$       & $0.57\pm 0.06$ & 181.72 &  1.04 &  22.19 &   1.31   \\
YZ Cet             & 3.72   & M4.5 & 3105 &  3052 & 2.810 &  $6.42\pm 0.02$       & $0.60\pm 0.06$ &  31.38 &  1.07 &   3.97 &   2.27   \\
TZ Ari             & 4.45   & M4.5 & 3105 &  2995 & 2.810 &  $6.65\pm 0.02$       & $0.62\pm 0.06$ &  26.11 &  1.06 &   3.27 & \nodata  \\
DY Eri C           & 5.03   & M4.5 & 3105 &  3058 & 2.810 &  $5.96\pm 0.02$       & $0.55\pm 0.06$ &  47.80 &  1.03 &   5.77 & $< 1.0$  \\
V577 Mon A         & 4.09   & M4.5 & 3105 &  3074 & 2.810 &  $5.49\pm 0.02$       & $0.59\pm 0.06$ &  73.14 &  1.07 &   9.17 &   1.33   \\
LHS 451            & 5.01   & M4.5 & 3105 &  3302 & 2.710 &  $5.61\pm 0.02$       & $0.46\pm 0.06$ &  59.43 &  1.05 &   7.28 & \nodata  \\
EZ Aqr             & 3.45   & M5   & 3030 &  2855 & 2.943 &  $5.54\pm 0.02$       & $0.65\pm 0.06$ &  78.21 &  1.01 &   9.31 & $< 1.1$  \\
LHS 39             & 4.85   & M5.5 & 2940 &  2811 & 2.943 &  $7.84\pm 0.03$       & $0.69\pm 0.06$ &   9.65 &  1.03 &   1.17 & $< 9.0$  \\
FL Vir             & 4.39   & M5.5 & 2940 &  2866 & 2.943 &  $6.04\pm 0.02$       & $0.63\pm 0.06$ &  49.04 &  1.00 &   5.80 & $< 1.0$  \\
Proxima Cen        & 1.30   & M5.5 & 2940 &  2855 & 2.943 &  $4.35\pm 0.02$\tnm{a}& $0.64\pm 0.06$ & 234.04 &  1.00 &  27.80 & \nodata  \\
Ross 248           & 3.16   & M5.5 & 2940 &  2928 & 2.940 &  $5.93\pm 0.02$       & $0.60\pm 0.06$ &  52.46 &  1.01 &   6.23 &   1.72   \\
LHS 2              & 4.69   & M5.5 & 2940 &  2792 & 2.943 &  $7.44\pm 0.02$       & $0.72\pm 0.06$ &  14.10 &  1.04 &   1.73 & $< 4.5$  \\
LHS 1565           & 3.69   & M5.5 & 2940 &  2879 & 2.940 &  $6.61\pm 0.02$       & $0.66\pm 0.06$ &  28.80 &  1.03 &   3.50 & \nodata  \\
V1581 Cyg A        & 4.54   & M5.5 & 2940 &  2756 & 2.940 &  $6.85\pm 0.02$       & $0.79\pm 0.08$ &  24.82 &  1.09 &   3.20 & $<10.3$  \\
LHS 1375           & 8.50   & M5.5 & 2940 &  2821 & 3.034 &  $8.98\pm 0.05$       & $0.65\pm 0.09$ &   3.36 &  0.99 &        &          \\
LHS 288            & 4.49   & M5.5 & 2940 &  2832 & 2.943 &  $7.73\pm 0.03$       & $0.67\pm 0.06$ &  10.54 &  1.02 &   1.27 & \nodata  \\
UV/BL Cet          & 2.68   & M6   & 2850 &  2874 & 2.940 &  $5.34\pm 0.02$       & $0.66\pm 0.06$ &  93.03 &  1.03 &  11.32 & $< 0.5$  \\
GL 316.1           & 14.06  & M6   & 2850 &  2724 & 3.034 & $10.05\pm 0.05$       & $0.74\pm 0.12$ &   1.33 &  1.02 &        &          \\
Wolf 359           & 2.39   & M6   & 2850 &  2857 & 2.943 &  $6.08\pm 0.02$       & $0.69\pm 0.06$ &  47.51 &  1.05 &\nodata & \nodata  \\
V1581 Cyg B        & 4.54   & M6   & 2850 &  2602 & 3.030 &  $7.39\pm 0.02$       & $0.88\pm 0.07$ &  16.89 &  1.07 &   2.14 & $<15.4$  \\
GJ 283B            & 8.90   & M6.5 & 2750 &\tnm{e}&\nodata&  $9.26\pm 0.05$       &$-0.32\pm 0.15$ &   2.48 &  0.42\tnm{e} &        &          \\
DX Cnc             & 3.63   & M6.5 & 2750 &  2746 & 3.030 &  $7.26\pm 0.02$       & $0.71\pm 0.06$ &  17.12 &  1.00 &   2.04 & $< 4.4$  \\
LHS 292            & 4.54   & M6.5 & 2750 &  2675 & 3.069 &  $7.93\pm 0.03$       & $0.73\pm 0.06$ &   9.67 &  0.97 &   1.12 & $< 5.1$  \\
LHS 2930           & 9.63   & M6.5 & 2750 &  2711 & 3.069 &  $9.79\pm 0.05$       & $0.69\pm 0.10$ &   1.70 &  0.96 &        &          \\
GJ 644C            & 6.45   & M7   & 2650 &  2700 & 3.060 &  $8.82\pm 0.05$       & $0.74\pm 0.08$ &   4.19 &  0.99 &        &          \\
LHS 3003           & 6.40   & M7   & 2650 &  2704 & 3.060 &  $8.93\pm 0.05$       & $0.73\pm 0.08$ &   3.78 &  0.99 &        &          \\
2MASS J0952-1924   &\nodata & M7   & 2650 &\tnm{e}&\nodata& $10.87\pm 0.05$       & $0.55\pm 0.16$ &   0.67 &  0.79\tnm{e} &        &          \\
2MASS J2306-0502   &\nodata & M7.5 & 2550 &  2479 & 3.070 & $10.30\pm 0.05$       & $0.99\pm 0.12$ &   1.26 &  1.07 &        &          \\
LP 044-162         & 12.5   & M7.5 & 2750 &  2589 & 3.070 & $10.40\pm 0.05$       & $0.85\pm 0.11$ &   1.06 &  1.02 &        &          \\
LP 349-025         & 7.8    & M8   & 2450 &  2570 & 3.100 &  $9.57\pm 0.05$       & $0.84\pm 0.09$ &   2.30 &  1.00 &        &          \\
LP 326-021         &\nodata & M8   & 2450 &  2559 & 3.100 & $10.62\pm 0.05$       & $0.86\pm 0.13$ &   0.88 &  1.01 &        &          \\
LP 412-031         & 14.64  & M8   & 2450 &  2728 & 3.056 & $10.64\pm 0.05$       & $0.68\pm 0.13$ &   0.77 &  0.96 &        &          \\
LP 771-021         & 16.23  & M8   & 2450 &  2591 & 3.100 & $11.42\pm 0.05$       & $0.82\pm 0.15$ &   0.41 &  0.99 &        &          \\
LHS 2397 A         & 14.29  & M8   & 2450 &  2380 & 3.100 & $10.74\pm 0.05$       & $1.09\pm 0.10$ &   0.92 &  1.08 &        &          \\
ESO 207-61         & 18.48  & M8   & 2450 &\tnm{e}&\nodata& $12.11\pm 0.05$       & $0.49\pm 0.28$ &   0.25 &  0.65\tnm{e} &        &          \\
TVLM 513-46546     & 9.82   & M8.5 & 2350 &  2404 & 3.120 & $10.71\pm 0.05$       & $1.11\pm 0.13$ &   0.92 &  1.12 &        &          \\
LP 944-20          & 4.97   & M9   & 2250 &  2123 & 3.150 &  $9.55\pm 0.02$       & $1.41\pm 0.06$ &   3.58 &  1.10 &   0.48 & $<18.8$  \\
LHS 2065           & 8.53   & M9   & 2250 &  2420 & 3.150 &  $9.94\pm 0.05$       & $1.00\pm 0.09$ &   1.85 &  1.03 &        &          \\
TVLM 868-110639    & 17.39  & M9   & 2250 &  2193 & 3.150 & $11.35\pm 0.05$       & $1.30\pm 0.09$ &   0.63 &  1.08 &        &          \\
BRI 0021-0214      & 12.12  & M9.5 & 2150 &  2284 & 3.180 & $10.54\pm 0.05$       & $1.16\pm 0.08$ &   1.21 &  1.04 &        &          \\
2MASSW J1733+4633  &\nodata & M9.5 & 2150 &  2512 & 3.180 & $11.89\pm 0.05$       & $0.87\pm 0.14$ &   0.28 &  0.99 &        &          \\
DENIS-P J0021.0-4244&\nodata & M9.5 & 2150 &\tnm{d}&\nodata& $12.30\pm 0.05$       & $1.20\pm 0.18$ &   0.22 &  1.21\tnm{d} &        &          \\
\enddata

\tnt{*}{K$_S$ is the 2MASS magnitude. K magnitudes footnoted $^a$, $^b$ and $^c$ were converted to K$_S$ using the 
        2MASS-CIT transformation in \citet{cut03b}}
\tnt{\dag}{24um zero point magnitude taken as 7.14 Jy.}
\tnt{\ddag}{Predicted values of the flux at 24 and 70 \micron, P$_{24}$ and P$_{70}$ are derived as described in the text, 
except that footnoted values use T$_{eff}$ instead of T$_{ir\kern-0.1em fm}$ Upper limits are at the 2 $\sigma$ level.}
\tnt{a}{Photometry from Mould, J.R. \& Hyland, A.R., 1976, \apj, 308, 399} 
\tnt{b}{Photometry from Persson et al., 1977, \aj, 82, 9}
\tnt{c}{Photometry from Leggett, S.K. 1992, \apjs, 82, 351}
\tnt{d}{Excluded from K$_S$-[24] calculation due to low SNR, but fits the trend.}
\tnt{e}{Excluded from K$_S$-[24] calculation due to low SNR, and is highly discrepant.}

\end{deluxetable}
\clearpage

\clearpage

\clearpage

\clearpage

\begin{thebibliography}{}

\bibitem[Allard, et al.(2001)]{all01} Allard, F., Hauschildt, P., Alexander, D. R., Tamanai, A., \& Schweitzer, A. 2001, \apj, 556, 357
\bibitem[Aumann, et al.(1984)]{au84} Aumann, H. H., et al. 1984, \apj, 278, L23
\bibitem[Backman, et al.(1986)]{back86} Backman, D. E., Gillett, F. C. \& Low, F. J. 1986, Advances in Space Research, 6, 43.
\bibitem[Beichman, et al.(2005a)]{bei05a} Beichman, C. A., et al. 2005a, \apj, 626, 1061
\bibitem[Beichman, et al.(2005b)]{bei05b} Beichman, C. A., etal. 2005b, \apj, 622, 1160
\bibitem[Bryden, et al.(2006)]{bry05} Bryden, G., et al. 2006, \apj, 636, 1098
\bibitem[Blackwell, et al.(1986)]{blac86} Blackwell, et al. 1986, \mnras, 221, 427
\bibitem[Brott \& Hauschildt(2005)]{brot05} Brott, I., \& Hauschildt, P. H. 2005, ``The Three-dimensional 
Universe with GAIA", (ESA SP-576), ed. C. Turon, K. E. O'Flaherty, M. A. C. Perryman, p. 565.
\bibitem[Burgasser, et al.(2002)]{bur02} Burgasser, A. T., et al. 2002, \apj, 564, 421
\bibitem[Butler, et al.(2004)]{but04} Butler, R. P., et al. 2004, \apj, 617, 580
\bibitem[Chen, et al.(2005)]{chen05} Chen, C., et al. 2005, \apj, 634, 1372
\bibitem[Cruz, et al.(2003)]{cruz03} Cruz K. L., et al. 2003, \aj, 126, 2421
\bibitem[Cutri, et al.(2003a)]{cut03a} Cutri, R. M., et al. 2003a,``The 2MASS All-Sky Catalog of Point Sources'', University of Massachusetts and Infrared Processing and Analysis Center (IPAC/California Institute of Technology); VizieR On-line Data Catalog: II/246
\bibitem[Cutri, et al.(2003b)]{cut03b} Cutri, R. M., et al. 2003b. ``The 2MASS All-Sky Data Release Explanatory 
Supplement: VI.4 2MASS Photometric System'', University of Massachusetts and Infrared Processing and 
Analysis Center (IPAC/California Institute of Technology)
\bibitem[Dawson \& de Robertis(2000)]{dr00} Dawson, P. C., and de Robertis, M. M. 2000, \aj, 120, 1532
\bibitem[Dole, et al.(2004)]{dol04} Dole, H. et al. 2004 \apjs, 154, 93
\bibitem[Engelbracht, et al.(2007)]{eng06} Engelbracht, et al. 2007 \pasp, in press
\bibitem[Dullemond \& Dominik(2005)]{dd05} Dullemond, C. P., \& Dominik, C. 2005 \aap, 434, 971
\bibitem[Gizis, et al.(2000)]{giz00} Gizis, J. E., et al. 2000 \aj, 120, 1085
\bibitem[Goldreich, et al.(2004)]{gold04} Goldreich, P., Lithwick, Y., \& Sari, R. 2004, \apj, 614, 497
\bibitem[Golimowski, et al.(2004)]{golim04} Golimowski, D. A., Leggett, S. K., Marley, M. S., Fan, X., Geballe, T. R., Knapp, G. R., Vrba, F. J. et al. 2004, \aj, 127, 3516
\bibitem[Gordon, et al.(2004)]{gord04} Gordon, K. D., et al. 2004, \procspie, 5487, 177
\bibitem[Gordon, et al.(2005)]{gord05} Gordon, K. D., et al. 2005, \pasp, 117, 503
\bibitem[Henry, et al.(1999)]{hen99} Henry, T. J., et al. 1999, \apj, 512, 864
\bibitem[Johnson (1983)]{jon83} Johnson, H. M. 1983, \apj, 273, 702
\bibitem[Jura, et al.(2004)]{jur04} Jura, M., et al. 2004, \apjs, 154, 453
\bibitem[Kalas, et al.(2004)]{kal04} Kalas, P., Liu, M. C. \& Matthews, B.C. 2004, Science, 303, 1990
\bibitem[Kenyon \& Bromley(2004)]{kb04} Kenyon, S. J., \& Bromley, B. C. 2004, \aj, 127, 513
\bibitem[Kim, et al.(2005)]{kim05} Kim, J. S., et al. 2005, \apj, 632, 659
\bibitem[Krist(2005)]{krist05} Krist, J. E. 2005, http://ssc.spitzer.caltech.edu/archanaly/contributed/stinytim.tar.gz
\bibitem[Lang(1991)]{lang91} Lang, K. R. 1991, Astrophysical Data, Springer-Verlag, New York
\bibitem[Leinhardt \& Richardson(2005)]{lr05} Leinhardt, Z. M., \& Richardson, D. C. 2005, \apj, 625, 427
\bibitem[Liu, et al.(2004)]{liu04} Liu, M.C., Matthews, B.C., Williams, J. P., and Kalas, P. G. 2004, \apj, 608, L526
\bibitem[Lestrade, et al.(2006)]{les06} Lestrade, J. F., Wyatt, M. C., Bertoldi, F., Dent, W. R. F. \& Menten, K. M. 2006, \aap, 460, 733
\bibitem[Low, et al.(2005)]{low05} Low, F. J., Smith, P. S., Werner, M. W., Chen, C., Krause, K., Jura, M., and 
Hines, D. C. 2005, \apj, 631, 1170
\bibitem[Mould \& Hyland(1976)]{mh76} Mould, J. R. \& Hyland, A. R. 1976, \apj, 208, 399
\bibitem[Patten, et al. (2006)]{pat06} Patten, B. M., et al. 2006, \apj, 651, 502
\bibitem[Persson, et al.(1977)]{per77} Persson, S. E., Aaronson, M. \& Frogel, J. A. 1977, \aj, 82, 729
\bibitem[Plavchan, et al.(2005)]{plav05} Plavchan, P., Jura, M. \& Lipscy, S. J. 2005, \apj, 631, 1161
\bibitem[Reid \& Hawley(2000)]{rh00} Reid, N. \& Hawley, S. L. 2000, ``New Light on Dark Stars," (Springer: New York)
\bibitem[Ribas(2003)]{rib03}Ribas, I. 2003, \aap, 400, 297
\bibitem[Rieke, et al.(2004)]{gr04} Rieke, G. H., et al. 2004, \apjs, 154, 25
\bibitem[Rieke, et al.(2005)]{gr05} Rieke, G. H., et al. 2005, \apj, 620, 1010
\bibitem[Song, et al.(2002)]{song02} Song, I., Weinberger, A. J., Becklin, E. E., Zuckerman, B. \& Chen, C. 2002, \aj, 124, 514
\bibitem[Stansberry, et al.(2007)]{stan07}Stansberry, J., et al. 2007, \pasp, in press
\bibitem[Su, et al.(2006)]{su06} Su, K., et al. 2006, \apj, 653, 675
\bibitem[Tokunaga(2000)]{tok00} Tokunaga, A. T. 2000, in \textit{Allen's Astrophysical Quantities}, 4th 
edition, ed. A.N. Cox, Springer-Verlag, NY, p. 143
\bibitem[Werner, et al.(2004)]{wer04} Werner, M. W., et al. 2004, \apjs, 154, 1
\end{thebibliography}
\end{document}